\documentstyle[A4,epsfig]{article}
\begin{document}
\vspace*{-1.8cm}
\begin{flushright}
\flushright{\bf LAL/RT 00-08}\\
\vspace*{-0.2cm}
\flushright{October 2000}
\end{flushright}
\vskip 1. cm
\begin{center}
{\Large \bf An Improved Empirical Equation for Bunch\\Lengthening 
in Electron Storage Rings}\\
\vspace{8mm}
{\large \bf J. Gao}\\
\vspace{3mm}
{\bf Laboratoire de L'Acc\'el\'erateur Lin\'eaire,}\\
IN2P3-CNRS et Universit\'e de Paris-Sud, BP 34, 91898 Orsay cedex, France\\
\end{center}
\vspace{6mm}
\begin{abstract}
In this paper we propose an improved 
empirical equation for the bunch lengthening
in electron storage rings. The comparisons are made between
the analytical and experimental results, and the agreements are quite well.
This improved equation can be equally 
applied to the case where a storage ring is
very resistive (such as the improved SLC damping rings)
instead of inductive as usual.
\par
\end{abstract}
\baselineskip=17pt

\section{Introduction}
From what we know about the single bunch longitudinal and transverse
instabilities \cite{Gao1}\cite{Gao2}, it is clear to see that the information 
about the bunch lengthening, ${\bf R}_z=\sigma_z/\sigma_{z0}$, with respect
to the bunch current is the {\it key} to open the locked chain of
bunch lengthening, energy spread increasing 
and the fast transverse instability threshold current. 
In this paper an improved (compared with what we have proposed in ref. 3) 
empirical bunch lengthening equation is proposed as follows:  
\begin{equation}
{\bf R}_z^2=1+{\sqrt{2{\cal C}}R_{av}R{\cal D}{\cal K}_{||,0}^{\rm tot}I_b
\over  \gamma^{3.5}({\bf R}_z)^{\varsigma}}+
{{\cal C}(R_{av}RI_b{\cal D}{\cal K}_{||,0}^{\rm tot})^2
\over \gamma^7({\bf R}_z)^{2\varsigma}}
\label{eq:gj11}
\end{equation}
where
\begin{equation}
{\cal C}={576\pi^2\epsilon_0\over 55\sqrt{3}\hbar c^3}
\label{eq:gj9}
\end{equation}
$${\cal D}=Exp\left(-\left({10\over 2\pi}arctan\left(
{Z_r\over Z_i}\right)\right)^2\right)$$
\begin{equation}
=Exp\left(-\left({10\over 2\pi}arctan\left(
{{\cal K}_{||,0}^{\rm tot}\over 2\pi {\cal L}}\left({3\sigma_{z0}\over c}
\right)^2\right)\right)^2\right)
\label{eq:gj9a}
\end{equation}
$\sigma_{z0}$ is the single particle "bunch length", 
${\cal K}_{||,0}^{\rm tot}$ is the
bunch total longitudinal loss factor for one turn at $\sigma_{z}=\sigma_{z0}$,
$Z_r$ and $Z_i$ are the resistive and inductive part of the machine impedance, 
respectively, 
${\cal L}$ is the inductance of the ring for one turn,
$\epsilon_0$ is the permittivity in vacuum, $\hbar$ is Planck constant, $c$
is the velocity of light,
$I_b=eN_ec/2\pi R_{av}$, $N_e$ is the particle number inside the bunch, 
and $R_{av}$ is the average radius of the ring.
Obviously, if $Z_i \gg Z_r$ one has ${\cal D} \approx 1$, which is the case
for the most existing storage rings.
If SPEAR scaling law \cite{PW} is used (for example), $\varsigma \approx 1.21$
(in fact 
each machine has its own $\varsigma$), eq. \ref{eq:gj11} can be written as
\begin{equation}
{\bf R}_z^2=1+{\sqrt{2{\cal C}}R_{av}R{\cal D}{\cal K}_{||,0}^{\rm tot}I_b
\over  \gamma^{3.5}{\bf R}_z^{1.21}}+
{{\cal C}(R_{av}RI_b{\cal D}{\cal K}_{||,0}^{\rm tot})^2
\over \gamma^7{\bf R}_z^{2.42}}
\label{eq:gj12}
\end{equation}
In fact, the third term of eqs. \ref{eq:gj11} is due to the {\it Collective 
Random Excitation} effect revealed in ref. 1, except a new factor ${\cal D}$
which is introduced in this paper to include the special case where $Z_i$ has
the same order of magnitude or even less than $Z_r$. 
The second term, however, is obtained intuitively as explained in section 3. 
Now we make more discussions on $Z_i$ and $Z_r$. Being aware of the possible
ambiguity coming from this frequently used term in the domain of collective
instabilities in storage rings, we define $Z_r$ and $Z_i$ used in this paper
as follows:
\begin{equation}
Z_r={P_b\over I_b^2}={{\cal K}_{||,0}^{\rm tot}T_b^2\over T_0}
\label{eq:gj12cc}
\end{equation}
and 
\begin{equation}
Z_i={2\pi\over T_0}{\cal L}
\label{eq:gj12dd}
\end{equation}
where $P_b=e^2N_e^2{\cal K}_{||,0}^{\rm tot}/T_0$, $I_b=eN_e/T_b$, 
$T_b=3\sigma_{z0}/c$, and $T_0$ is the particle revolution period.
By using eqs. \ref{eq:gj12cc} and \ref{eq:gj12dd} one gets explicit expression
of ${\cal D}$ shown in eq. \ref{eq:gj9a}.
\par
The procedure to get the information about the bunch lengthening and the 
energy spread increasing is firstly to find 
${\bf R}_z(I_b)$ by solving bunch lengthening equation, i.e., 
eq. \ref{eq:gj11}, 
and then calculate energy spread increasing, ${\bf R}_{\varepsilon}(I_b)$ 
(${\bf R}_{\varepsilon}=\sigma_{\varepsilon}/\sigma_{\varepsilon,0}$), 
by putting ${\bf R}_z(I_b)$ into eq. \ref{eq:gj12a} \cite{Gao1}: 
\begin{equation}
{\bf R}_{\varepsilon}^2=1+{{\cal C}(R_{av}RI_b{\cal D}
{\cal K}_{||,0}^{\rm tot})^2
\over \gamma^7{\bf R}_z^{2.42}}
\label{eq:gj12a}
\end{equation}
Once ${\bf R}_{\varepsilon}(I_b)$
is found, one can use the following formula 
to calculate the fast single bunch transverse
instability threshold current \cite{Gao2}:
\begin{equation}
I^{th}_{b,gao}={F'f_sE_0\over e<\beta_{y,c}>
{\cal K}^{\rm tot}_{\perp}(\sigma_z)}
\label{eq:30a}
\end{equation}
with
\begin{equation}
F'=4{\bf R}_{\varepsilon}
\vert \xi_{c,y}\vert
{\nu_y\sigma_{\varepsilon 0}\over \nu_sE_0}
\label{eq:30b}
\end{equation}
where $\nu_s$ and $\nu_y$ are synchrotron 
and vertical betatron oscillation tunes,
respectively, $<\beta_{y,c}>$ is the average beta
function in the rf cavity region, $\xi_{c,y}$ is the
chromaticity in the vertical plane (usually positive to
control the head-tail instability),
${\cal K}^{\rm tot}_{\perp}(\sigma_z)$ is the total transverse loss factor
over one turn, $\sigma_{\varepsilon 0}$ is the natural energy spread, and 
$E_0$ is the particle energy. 
In practice, it is useful to express ${\cal K}^{\rm tot}_{\perp}(\sigma_z)$ as
${\cal K}^{\rm tot}_{\perp}(\sigma_z)={\cal K}^{\rm  tot}_{\perp,0}/{\bf R}_z^{\Theta}$,
where ${\cal K}^{\rm tot}_{\perp,0}$ is the value at the natural bunch length,
and ${\Theta}$ is a constant depending on the machine concerned.
As a Super-ACO scaling law, ${\Theta}$ can be taken as $2/3$ \cite{Brun}. 
Eq. \ref{eq:30a} is therefore expressed as:
\begin{equation}
I^{th}_{b,gao}={F'f_sE_0{\bf R}_z^{2/3}\over e<\beta_{y,c}>
{\cal K}^{\rm tot}_{\perp,0}}
\label{eq:30c}
\end{equation}
The notation $I^{th}_{b,gao}$ is used with the aim of distinguishing it
from the formula given by Zotter \cite{Zotter2}\cite{Zotter3}. 
\par
\section{Comparison with Experimental Results}
In this section we look at seven machines
with their parameters shown in table \ref{tab:1}.\\
\begin{table}[h]
\begin{center}
\begin{tabular}{|l|l|l|}
\hline
Machine&$R$ (m)&$R_{av}$ (m)\\
\hline
INFN-A&1.15&5\\
\hline
ACO&1.11&3.41\\
\hline
SACO&1.7&11.5\\
\hline
KEK-PF&8.66&29.8\\
\hline
SPEAR&12.7&37.3\\
\hline
BEPC&10.345&38.2\\
\hline
SLC Damping Ring&2.037&5.61\\
\hline
\end{tabular}
\end{center} 
\caption{The machine parameters.}
\label{tab:1}
\end{table}

The machine energy, natural bunch length and the corresponding longitudinal
loss factor are given in table \ref{tab:2}. \\

\begin{table}[h]
\begin{center}
\begin{tabular}{|l|l|l|l|}
\hline
Machine&$\gamma$
&$\sigma_{z0}$ (cm)&${\cal K}_{||,0}^{\rm tot}$ (V/pC)\\
\hline
INFN-A&998&3.57&0.39\\
\hline
ACO&467&21.7&0.525\\
\hline
SACO&1566&2.4&3.1\\
\hline
KEK-PF&3523&1.1&5.4\\
\hline
KEK-PF&4892&1.47&3.7\\
\hline
SPEAR&2935&1&5.2\\
\hline
BEPC&2544&1&9.6\\
\hline
BEPC&3953&2&3.82\\
\hline
SLC Damping Ring&2329&0.53&12\\
\hline
\end{tabular}
\end{center} 
\caption{The machine energy and the total loss factors.}
\label{tab:2}
\end{table}
\noindent
Concerning the loss factors, that of
INFN accumulator ring comes from ref. 6 and the others are obtained by fitting
the corresponding experimental results with the bunch lengthening equation 
given in ref. 1.
Figs. 1 to 10 show the comparison results between the analytical 
and the experimental \cite{PW}-\cite{16}
bunch lengthening values, and Fig. 11 shows the single bunch energy spread
increasing. It is obvious that this
improved empirical bunch lengthening equation is quite powerful.
Among the seven different storage rings, 
SLC new damping ring is the unique and the most 
interesting one since it is a very resistive ring \cite{14}, 
on the contrary, the other rings including SLC old damping ring
are quite inductive.
The inductances of the old and the new SLC damping rings are 33 nH and 6 nH,
respectively \cite{15}. By fitting the bunch lengthening experimental 
results, one finds that the loss factor ${\cal K}_{||,0}^{\rm tot}$ equals
12 V/pC at $\sigma_{z0}=0.53$ cm (this value is put in table 2), 
which agrees quite well with the experimentally measured
loss factor, 15 V/pC, at the same bunch length \cite{16}. 
From Fig. 11 one can see that the single bunch energy spread increasing
in SLC new damping ring is rather accurately predicted by eq. \ref{eq:gj12a}.
\par
\section{Discussion}
In fact eq. \ref{eq:gj11} can be obtained from the following equation
by truncating the Taylor expansion of the right hand side of eq. \ref{eq:gj72}
up to the second order.
\begin{equation}
{\bf R}_z^2=\exp\left({\sqrt{2{\cal C}}R_{av}R{\cal D}
{\cal K}_{||,0}^{\rm tot}I_b
\over  \gamma^{7/2}{\bf R}_z^{\varsigma}}\right)
\label{eq:gj72}
\end{equation}
From the point of view of aesthetics, eq. \ref{eq:gj72} is more
attractive (at least for the author).
Even if it doesn't work well itself,
this equation is instructive
for us to establish the second term in eq. \ref{eq:gj11}.
\par
\section{Conclusion}
In this paper we propose an improved empirical bunch lengthening equation
and compare the analytical results with the experimental results
of seven different machines where SLC new damping ring is quite
resistive. The agreement between the analytical and
experimental results is quite satisfactory. 
The factor ${\cal D}$ introduced in this paper should be
included (one should multiply it to ${\cal K}_{||,0}^{\rm tot}$)
into the corresponding formulae in ref. 1 also in order to 
be applied to the case where a storage ring is very resistive.
\par
\section{Acknowledgement}
The author thanks J. Le Duff and J. Ha\"\i ssinski 
for their critical comments and interests in this subject.
I have enjoyed the interesting discussions on SLC damping rings
with K. Bane, B. Podobedov, A. Chao, G. Stupakov, S. Heifets, and some other 
theory club members at SLAC.

\par
\begin{figure}[h]
\centering
\vskip -0. true cm
\mbox{\epsfig{figure=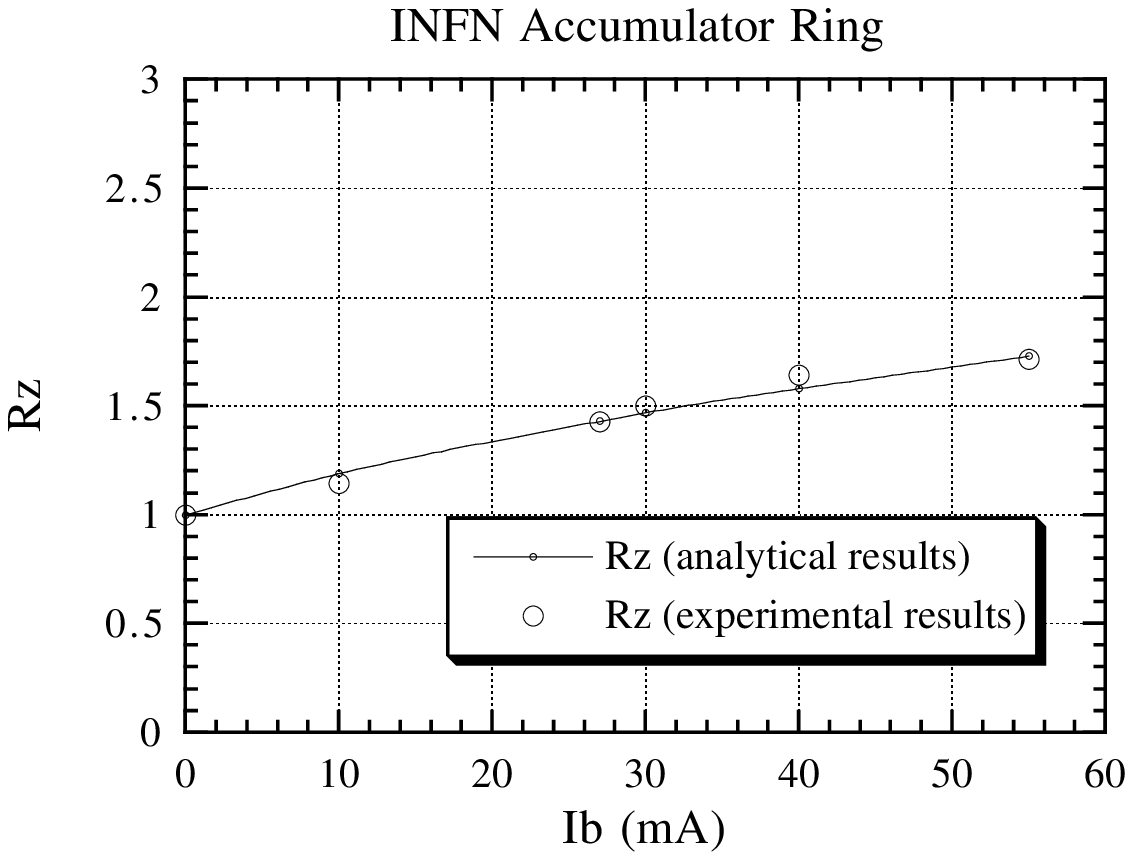, height=6.5cm,%
 width=8.54cm, bbllx=73pt, bblly=270pt, bburx=512pt, bbury=599pt}}
\vskip -0 true cm
\caption{Comparison between INFN accumulator ring ($R=1.15$ m and $R_{av}=5$ m)
experimental results
and the analytical results at 510 MeV with $\sigma_{z_0}$=3.57 cm.
\label{fig:1}}
\end{figure}
\begin{figure}[h]
\begin{center}
\vskip -1.8 true cm
\mbox{\epsfig{figure=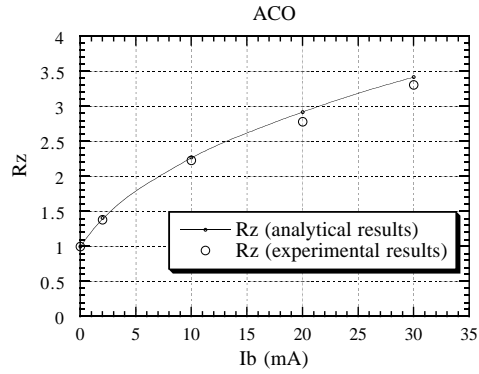,height=6.5cm,%
width=8.54cm,bbllx=73pt,bblly=270pt,bburx=512pt,bbury=599pt}}
\end{center}
\vskip -0 true cm
\caption{Comparison between ACO ($R=1.11$ m and $R_{av}=3.41$ m)
experimental results
and the analytical results at 238 MeV with $\sigma_{z_0}$=21.7 cm.
\label{fig:2}}
\end{figure}
\begin{figure}[h]
\begin{center}
\vspace*{-30mm}
\mbox{\epsfig{figure=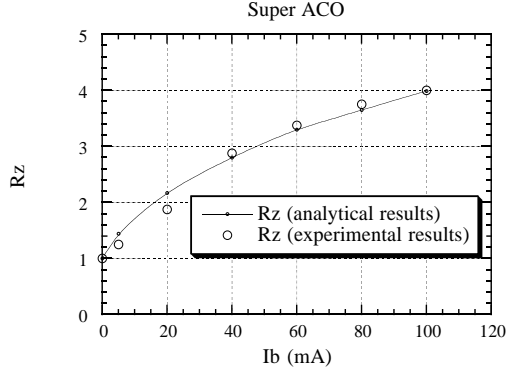,height=6.5cm,%
width=8.54cm,bbllx=73pt,bblly=270pt,bburx=512pt,bbury=599pt}}
\end{center}
\vskip -0. true cm
\caption{Comparison between Super-ACO ($R=1.7$ m and $R_{av}=11.5$ m)
experimental results
and the analytical results at 800 MeV with $\sigma_{z_0}$=2.4 cm.
\label{fig:3}}
\end{figure}
\begin{figure}[h]
\begin{center}
\vskip -1.6 true cm
\mbox{\epsfig{figure=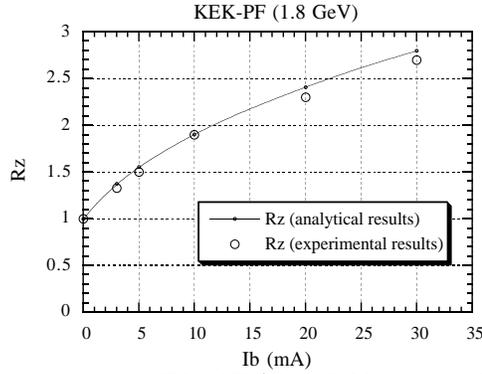,height=6.5cm,%
width=8.54cm,bbllx=73pt,bblly=270pt,bburx=512pt,bbury=599pt}}
\end{center}
\vskip -1 true cm
\caption{Comparison between KEK-PF ($R=8.66$ m and $R_{av}=29.8$ m)
experimental results
and the analytical results at 1.8 GeV with $\sigma_{z_0}$=1.47 cm.
\label{fig:4}}
\end{figure}
\begin{figure}[h]
\begin{center}
\vskip -1.5 true cm
\mbox{\epsfig{figure=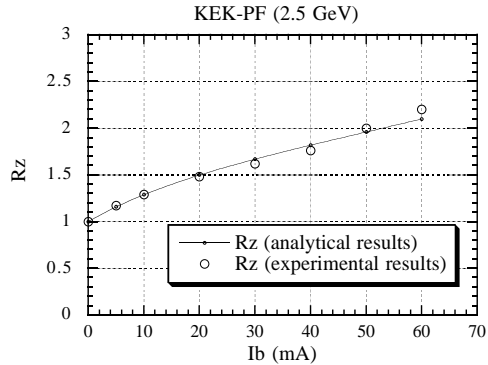,height=6.5cm,%
width=8.54cm,bbllx=73pt,bblly=270pt,bburx=512pt,bbury=599pt}}
\end{center}
\vskip -1 true cm
\caption{Comparison between KEK-PF (2.5 GeV) ($R=8.66$ m and  $R_{av}=29.8$ m)
experimental results
and the analytical results at 2.5 GeV with $\sigma_{z_0}$=1.1 cm.
\label{fig:5}}
\end{figure}
\newpage
\begin{figure}[h]
\begin{center}
\vskip -1.5 true cm
\mbox{\epsfig{figure=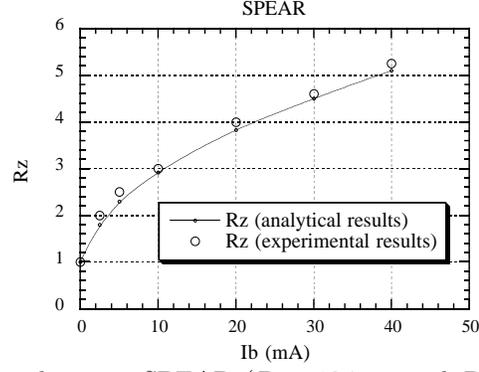,height=6.5cm,%
width=8.54cm,bbllx=73pt,bblly=270pt,bburx=512pt,bbury=599pt}}
\end{center}
\vskip -1 true cm
\caption{Comparison between SPEAR ($R=12.7$ m and $R_{av}=37.3$ m)
experimental results
and the analytical results at 1.5 GeV with $\sigma_{z_0}$=1 cm.
\label{fig:6}}
\end{figure}
\begin{figure}[h]
\begin{center}
\vskip -1.5 true cm
\mbox{\epsfig{figure=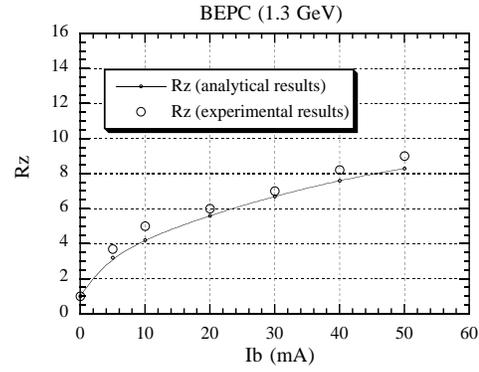, height=6.5cm,%
width=8.54cm,bbllx=73pt,bblly=270pt,bburx=512pt,bbury=599pt}}
\end{center}
\vskip -1 true cm
\caption{Comparison between BEPC (1.3 GeV) ($R=10.345$ m and $R_{av}=38.2$ m)experimental results and the analytical results at 1.3 GeV with $\sigma_{z_0}$=1 cm.
\label{fig:7}}
\end{figure}
\begin{figure}[h]
\begin{center}
\vskip -1.5 true cm
\mbox{\epsfig{figure=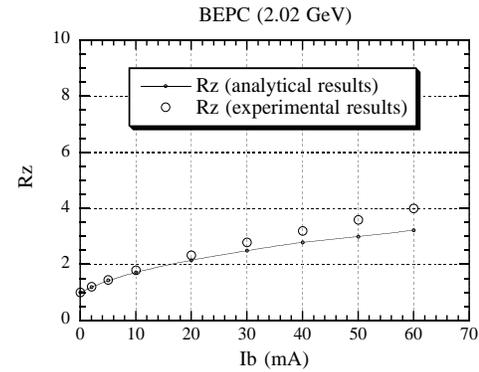, height=6.5cm,%
width=8.54cm,bbllx=73pt,bblly=270pt,bburx=512pt,bbury=599pt}}
\end{center}
\vskip -1 true cm
\caption{Comparison between BEPC ($R=10.345$ m and $R_{av}=38.2$ m)
experimental results
and the analytical results at 2.02 GeV with $\sigma_{z_0}$=2 cm.
\label{fig:8}}
\end{figure}

\begin{figure}[h]
\vskip -1.5 true cm
\vspace{6.0cm}
\includegraphics{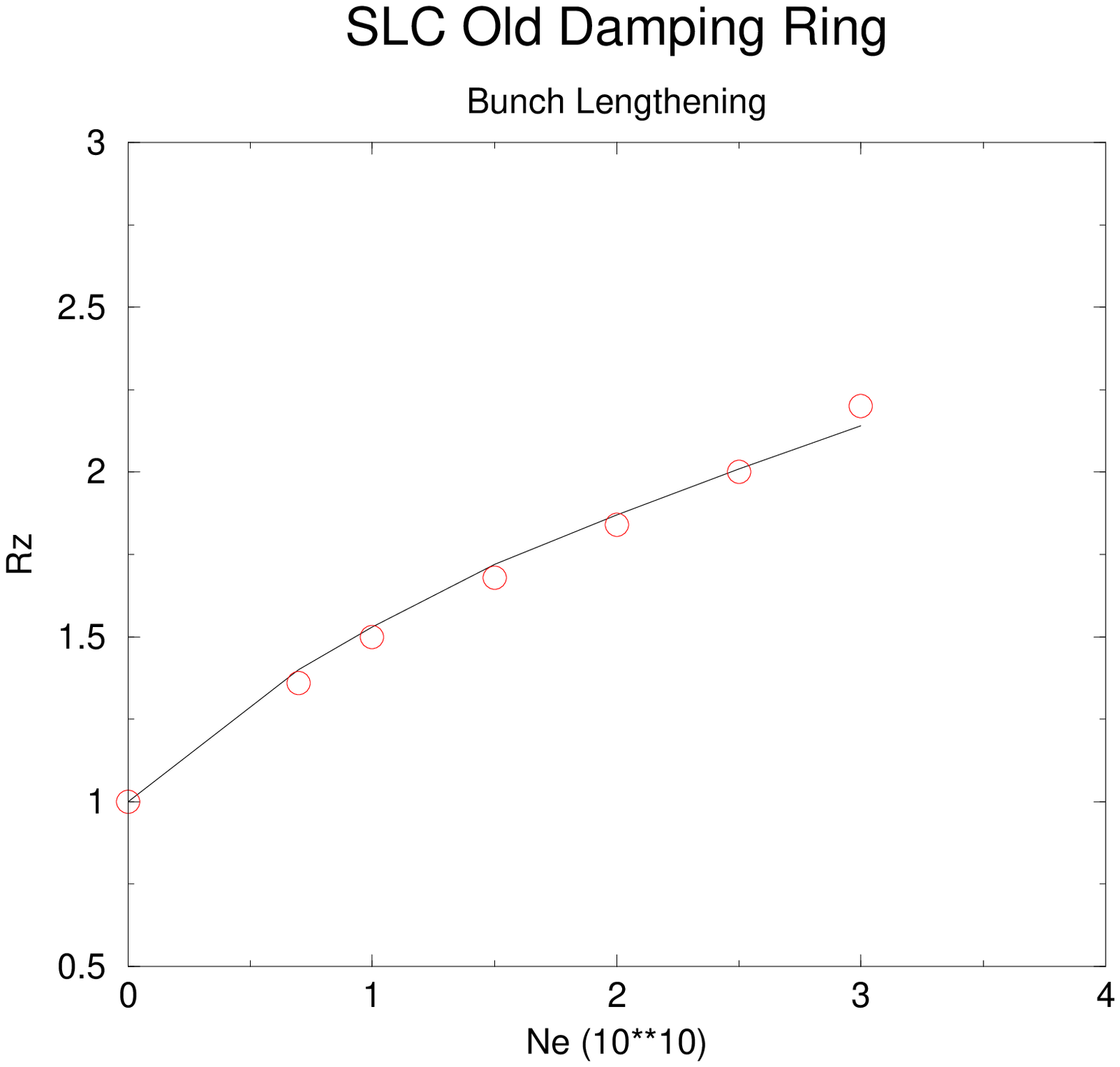}
\vskip  0 true cm
\caption{Comparison between SLC old damping ring ($R=2.037$ m 
and $R_{av}=5.61$ m)
experimental (circles)
and analytical (line) results of bunch lengthening
at 1.19 GeV with $\sigma_{z_0}$=0.53 cm.
\label{fig:8a}}
\end{figure}

\begin{figure}[h]
\vskip -1.5 true cm
\vspace{6.0cm}
\includegraphics{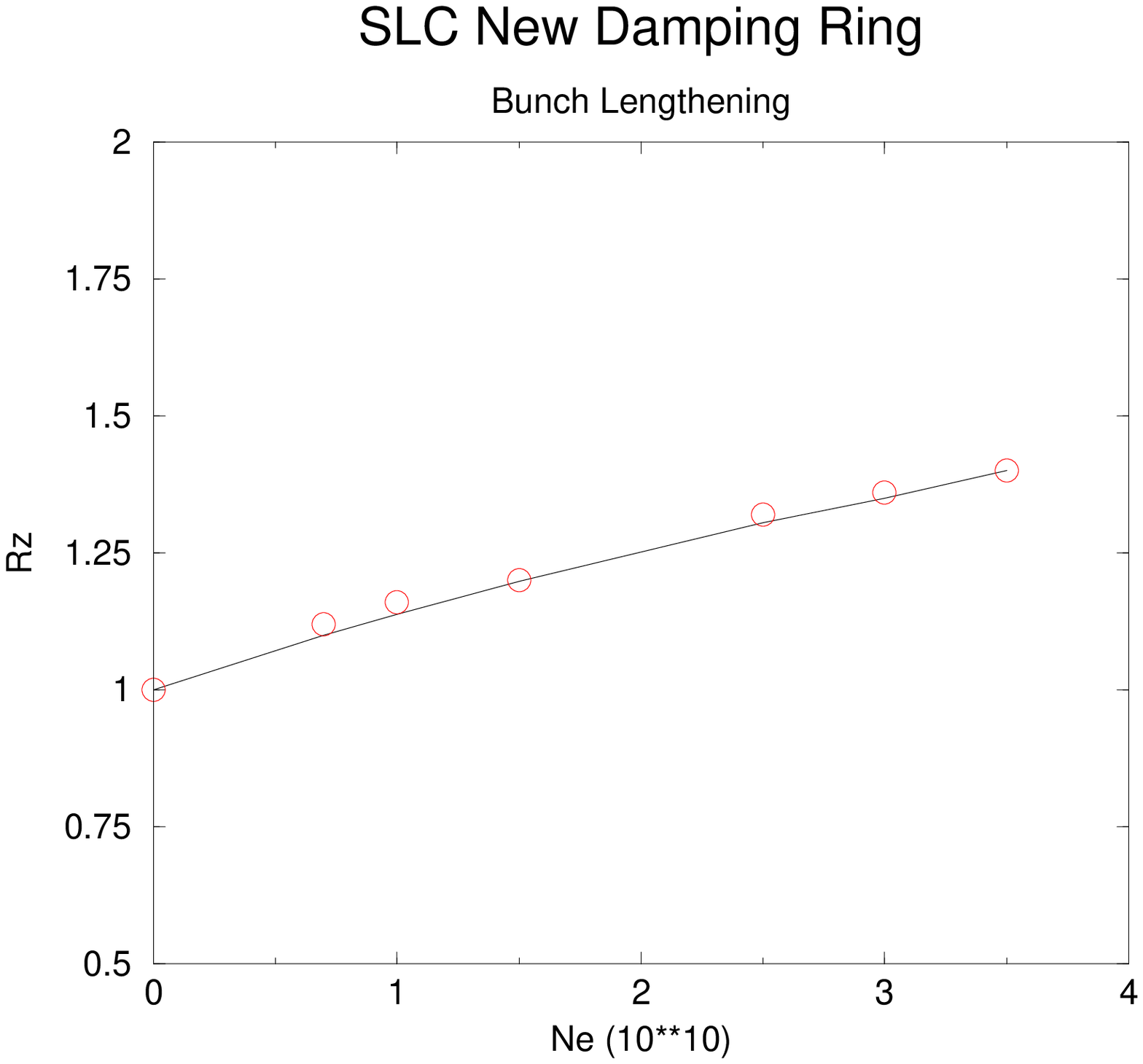}
\vskip  0 true cm
\caption{Comparison between SLC old damping ring ($R=2.037$ m
and $R_{av}=5.61$ m)
experimental (circles)
and analytica (line) results of bunch lengthening
at 1.19 GeV with $\sigma_{z_0}$=0.53 cm.
\label{fig:8b}}
\end{figure}
\begin{figure}[h]
\vskip -1.5 true cm
\vspace{6.0cm}
\includegraphics{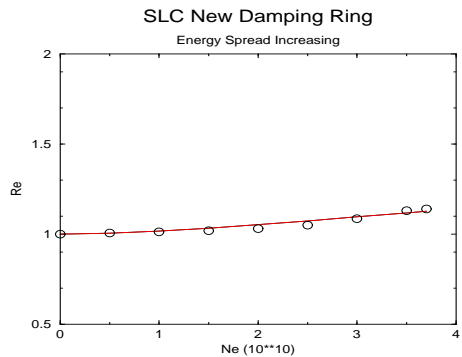}
\vskip  0 true cm
\caption{Comparison between SLC old damping ring ($R=2.037$ m
and $R_{av}=5.61$ m)
experimental (circles)
and analytical (line) results of energy spread increasing
at 1.19 GeV with $\sigma_{z_0}$=0.53 cm.
\label{fig:8c}}
\end{figure}

\end{document}